\documentclass[twocolumn,appendixfloats,tighten]{aastex6}
\usepackage{newtxtext,newtxmath}
\usepackage[T1]{fontenc}
\usepackage{ae,aecompl}
\usepackage{amsmath}	
\usepackage{amssymb}	
\usepackage{mathtools}
\usepackage{ctable}
\usepackage{url}
\usepackage{xspace}
\usepackage[normalem]{ulem}
\usepackage{hyperref}
\usepackage[all]{hypcap}
\usepackage{graphicx}
\usepackage{color}
\usepackage{float}
\usepackage{mathrsfs}

\def\msun{{\rm\,M_\odot}}

\def\msun{{\rm\,M_\odot}}

\def\msun{{\rm\,M_\odot}}

\newcommand{\be}{\begin{equation}}
\newcommand{\ee}{\end{equation}}

\newcommand{\tmin}{t_{\rm min}}

\def\h2{${\rm\,H_2}$}

\begin{document}

\title{constraining the delay time distribution of compact binary objects from the stochastic gravitational wave background searches}
\author{Mohammadtaher Safarzadeh\altaffilmark{1}, Sylvia Biscoveanu\altaffilmark{2,3}, Abraham Loeb\altaffilmark{1}}

\altaffiltext{1}{Center for Astrophysics | Harvard \& Smithsonian, 60 Garden Street, Cambridge, MA, USA
\href{mailto:msafarzadeh@cfa.harvard.edu}{msafarzadeh@cfa.harvard.edu}}
\altaffiltext{2}{LIGO Laboratory, Massachusetts Institute of Technology, 185 Albany St, Cambridge, MA 02139, USA}
\altaffiltext{3}{Department of Physics and Kavli Institute for Astrophysics and Space Research, MIT, 77 Massachusetts Ave, Cambridge, MA 02139, USA}

\begin{abstract}
The initial separation of massive star binaries sets the timescale over which their compact remnants merge through the emission of gravitational waves.
We show that the delay time distribution (DTD) of binary neutron stars or black holes can be inferred from the stochastic gravitational wave background (SGWB).
If the DTD of a population is long, most of the mergers take place at low redshifts and the background would be rather quiet compared to a scenario in which the DTD is short leading 
to few individual detections at low redshift but a rather loud background.
We show that different DTDs predict a factor of 5 difference in the magnitude of the gravitational wave background energy density ($\Omega_{\rm GW}$) 
and have the dominant effect on $\Omega_{\rm GW}$ over other factors such as the mass function of the 
primary BH mass, $p(m_1)$, the maximum considered BH mass ($M_{\rm max}$), and the effective spin of the black hole ($\chi_{\rm eff}$).
A non-detection of such a background can rule out the short DTD scenario. 
We show that SGWB searches can rule out the short DTD scenario for the BBHs within about four years of observing time at advanced LIGO design sensitivty for a local merger rate of 30 $\rm Gpc^{-3} yr^{-1}$ assuming 
$p(m_1)\propto m_1^{-1}$, and $M_{\rm max}=50 M_{\odot}$. 
\end{abstract}

\section{Introduction}

The detectability of individual compact binary objects (CBOs) through the emission of gravitational waves (GWs) by the advanced LIGO~\citep{Collaboration:2015fz} and Virgo~\citep{Acernese:2015kd} detectors depends on the CBO's mass, distance, and orientation on the sky.
While the redshift reach of LIGO to detected individual sources is $z\sim0.1$ for binary neutron stars (BNS), and $z\sim1$ for binary black holes (BBH), the majority of the coalescing binaries would be undetected 
as their Signal to Noise Ratio (SNR) falls below the detection threshold~\citep{Abbott:2018gf}. 
The undetected population of the CBOs contribute to the stochastic gravitational-wave background (SGWB) that is detectable through searches for excess correlated power in two or more GW detectors \citep{Allen:1997gg,Camp:2004km,Romano:2017fs}. 
The LIGO Scientific Collaboration currently constrains the dimensionless energy density of gravitational waves to be $\Omega_{\rm GW}<6.0 \times 10^{-8}$ with 95\% confidence, 
assuming a flat energy density spectrum in the most sensitive part of the LIGO band (20 -86 Hz) \citep{Abbottetal:2019kc}.

The contribution of each sub-population of the CBOs to the SGWB depends on the cosmic star formation history and their associated delay time distribution (DTD).
In this paper we use the known star formation history to examine the impact of the DTD on the GW background amplitude and frequency spectrum.
Assuming that the initial separations ($a$) of compact binaries follows $dN/da\propto a^{-1}$, given that the timescale for merging through gravitational waves scales as $t\propto a^4$, 
a power law distribution in coalescence timescales $dN/dt\propto t^{-1}$ would be inferred. This underlies the results from the population synthesis analysis of binary stellar evolution leading to the formation of BNS and BBH systems \citep{Dominik:2012cw}. 
However, current observations allow for a range of possible distributions that might contradict the above assumptions \citep{Beniamini:2016kwa,Safarzadeh:2018ub}.

One can characterize the shape of a DTD through two parameters: 
(i) the minimum delay time ($\tmin$) that translates into the smallest possible separation for two compact objects, and 
(ii) the slope ($\Gamma$) of the probability distribution in time $dN/dt\propto t^{\Gamma}$. This is the simplest case if we assume the population can be represented by a single DTD, meaning 
no bi-modality is present in the population and also assuming the DTD does not evolve with redshift. It is, however, plausible that the DTD is bimodal, evolves with redshift, or 
is determined by a different form than a power law, such as a log-normal distribution \citep{Simonetti:2019uq}.

In the context of BNS systems, three frameworks for constraining the DTD have been considered: 
(i) scaling relations between the host galaxies of BNS merger events \citep{Safarzadeh:2019dn}, requiring on the order of $\mathcal{O}(10^3)$ GW detection in the local universe, 
(ii) third generation gravitational wave detectors, such as Einstein Telescope (ET) and Cosmic Explorer (CE) \citep{Safarzadeh:2019kj} which was shown to take about one year of data acquisition, 
and (iii) detailed knowledge of the host galaxies of the BNS merger events \citep{Safarzadeh:2019dp}, which was shown to require $\mathcal{O}(10^2)$ detections. 
LIGO's horizon for detecting BNS mergers reaches out to $z\approx0.1$ which is insufficient to probe the redshift distribution of the BNS mergers and therefore constrain the 2D parametrized model for the DTD \citep{Safarzadeh:2019kj}. 

The same rationale applies to BBH mergers; however, in this case it is possible that the underlying DTD depends on the mass scale of the binaries and/or their effective spin distribution. \cite{Vitale:2019cy} showed that the delay time distribution and star formation rate can be measured with three months of observations of binary black hole mergers with third generation detectors. Evolution of the DTD with mass or spin can be incorporated into the framework of their analysis, which relies on hierarchical Bayesian inference.
For BNS systems, one typically assumes to have a fixed component mass of 1.4 $\msun$, an assumption that should be 
re-visited in light of the most recent LIGO detection of a massive BNS system, GW190425, with a total mass of $\approx 3.4 \msun$ \citep{Abbottetal:2020uq}.

In this \emph{Letter} we demonstrate how advanced LIGO (adLIGO) will be able to constrain the DTD through SGWB searches. 
In \S2 we compute the merger rate of the BNS and BBH systems given different assumptions regarding the DTD. In \S3 we measure the corresponding expected SGWB level from each system. 
In \S4 we present our results for adLIGO at design sensitivity as our detector network. In \S5 we discuss the role and prospects for future GW detectors, and in \S6 we summarize our work.

\section{Merger rate of the BNS and BBH systems with different DTDs}

\begin{figure}
 \includegraphics[width=\columnwidth]{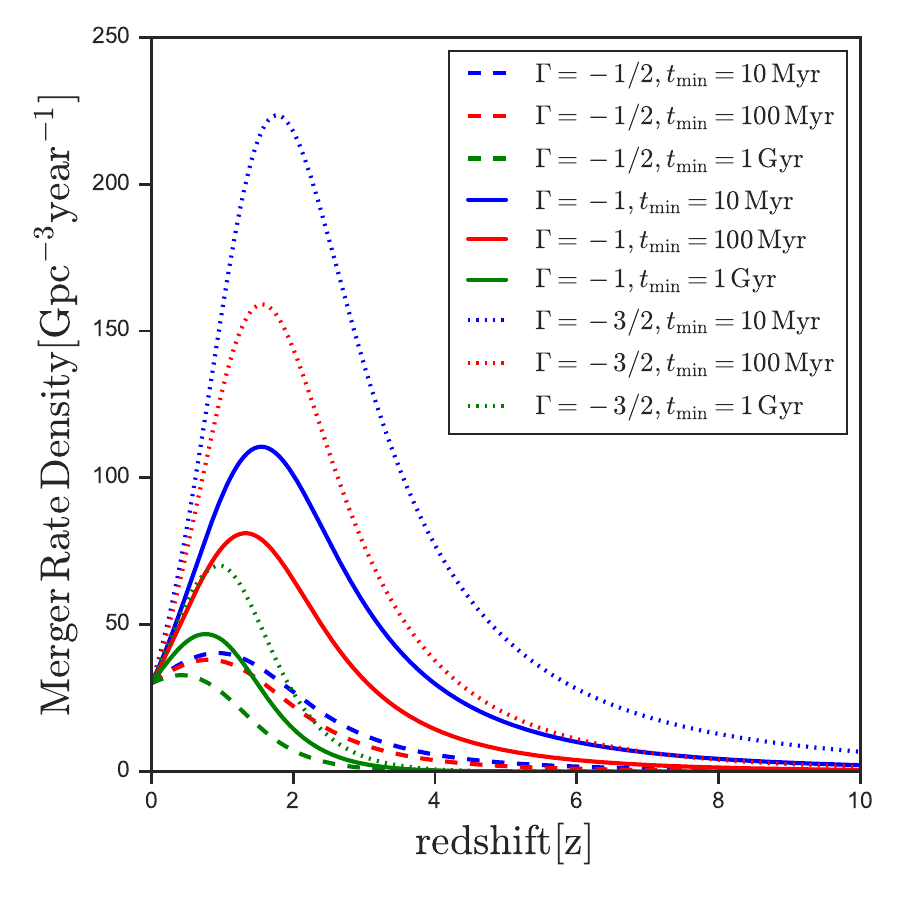}
 \caption{Merger rate history of BBH systems all normalized to 30 $ \rm Gpc^{-3}yr^{-1}$ at redshift $z=0$ for different DTDs.
 The DTDs favoring long delay, i.e., those with shallower slopes ($\Gamma=-1/2$) and longer minimum delay times ($\tmin=1 \rm Gyr$) merge a smaller fraction of all 
 binaries by $z=0$ compared to the DTD model that favors short delay times. }
 \label{fig:bbh_dtd}
\end{figure}

The merger rate as a function of redshift is a convolution of the DTD with the cosmic star formation rate density:
\begin{align}
R(z)=&\int_{z_b=10}^{z_b=z} \lambda\frac{dP_m}{dt}(t-t_b-t_{\rm min})\psi(z_b)\frac{dt}{dz}(z_b)dz_b,
\end{align}
where $dt/dz = -[(1+z) E(z) H_0]^{-1}$, and 
$E(z)=\sqrt{{\Omega}_{m,0}(1+z)^3+{\Omega}_{k,0}(1+z)^2+{\Omega}_{\Lambda}(z)}$. We use $H_{0} = 67~\mathrm{km\,s^{-1}\,Mpc^{-1}}$ for the Hubble constant, and $\Omega_{m,0} = 1-\Omega_{\Lambda} = 0.31$~\citep{Ade:2015xua}.
Here, $\lambda$ is the BNS or BBH production efficiency per mass in stars (assumed not to evolve with redshift) used as a free parameter to normalize the merger rate in the local universe; $t_b$ is the cosmic time corresponding to redshift $z_b$; $dP_m/dt$ is the DTD, parametrized to follow a power law distribution ($\propto t^{\Gamma}$) with a minimum delay time, $t_{\rm min}$ that refers to the time since the birth of the progenitor stars  
(the sum of the nuclear lifetime of the lowest mass component of the binary system and the minimal gravitational delay that is induced by the existence of a minimal separation between the two newly born compact objects). 
We also impose a maximum delay time of 10 Gyr for our fiducial case, comparable to the age of the universe. We choose to integrate from $z_{b}=10$, since mergers beyond this redshift are extremely unlikely, even for the fastest DTD models (see Figure~\ref{fig:bbh_dtd}). We adopt the cosmic star formation rate density from \citet{Madau:2014gtb}:
\be
\psi(z)=0.015 \frac{(1+z)^{2.7}}{1+[(1+z)/2.9]^{5.6}}\,\, \msun\, {\rm yr^{-1}\, Mpc^{-3}}.
\ee 

In this work we ignore the impact of metallicity with uncertainties that may affect the final merger rate of the BBHs, and BNSs. We refer the reader to other works that have explored the impact of such uncertainties \citep[e.g., ][]{Safarzadeh:2019fh,Neijssel:2019ca}. While in the approach presented in this work an overall sense of DTD could be constrained, in reality DTD of BBHs would depend on their mass as more massive BBHs are born preferentially at lower metallicities. 
However, DTD uncertainties likely dominate over that of star formation history (SFH) and metallicity evolution: For example, by fixing the SFH parametrization, and adopting different relations for the stellar mass-metallicity relation, \citet{Neijssel:2019ca} arrives at about less than 1 dex difference in the predicted merger rate of the BBHs (see their Table 1). Although relying on extreme uncertainties for SFH at high redshift can inflate the uncertainty budget for the merger rate of the BBHs, reasonable SFH parameterizations only differ by about 10-20\% in the overall shape and magnitude. We only caution the reader of other contributions to the merger rate uncertainties and defer a more comprehensive work to a future study.

Figure \ref{fig:bbh_dtd} shows the expected merger rate of BBHs across redshift for nine different DTD models all normalized to  30 $ \rm Gpc^{-3}yr^{-1}$ at $z=0$, which is a conservative estimate based on the inferred local merger rate from LIGO's second observing run~\citep{Abbottetal:2018vb}. These values of $\Gamma$ were chosen to represent small deviations from the expected value of $\Gamma=-1$ based on binary population synthesis~\citep{Dominik:2012cw}. 
In both cases, those DTDs favoring long delay, i.e., those with shallower slopes ($\Gamma=-1/2$) and longer minimum delay times ($\tmin=1 \rm Gyr$) merge a smaller fraction of all binaries by $z=0$ compared to the DTD model that favors short delay times.
This significant change in the expected merger rate at high redshifts leads to DTDs with shorter delay times having a significantly larger contribution to the SGWB that models in which the DTD favors long delays.

\section{The stochastic background}

The stochastic background is defined as the energy density of gravitational waves $\rho_{GW}$ per logarithmic frequency interval:
\be
\Omega(f)=\frac{1}{\rho_c}\frac{d~\rho_{GW}}{d~ln~f},
\ee
where $\rho_c$ is the critical density of the universe given by $\rho_c=3H_0^2/8\pi G$. Here $H_0$ is the Hubble constant, and $G$ is the Newton constant. 
The spectrum of the SGWB is given by:
 
\begin{equation}
\frac{d~\rho_{GW}}{d~ln~f}=
\int _{z=0}^{z=\infty}
\frac{R}{1+z}\frac{dt}{dz}\left(f_r\frac{dE_{\rm gw}}{df_r}\right) \Big| _{f_r=f(1+z)}dz
\label{eq:omega_gw_2}
\end{equation}
\citep{Phinney:2001ue}, where $f$ and $f_r=f(1+z)$ are the observed and rest frame GW frequencies, respectively. 

The GW spectrum from a coalescing BNS or BBH is given by
\begin{equation}
\frac{dE_{\rm gw}}{df_r}=\frac{(\pi G)^{2/3}M_{\rm chirp}^{5/3}}{3}\\
\begin{cases}
f_r^{-1/3}\mathscr{F_{\rm PN}}~~~~&f_r<f_1,\\[2pt]
\omega_{\rm m} f_r^{2/3}\mathscr{G_{\rm PN}}~~&f_1\leq f_r<f_2,\\[2pt]
\dfrac{\omega_{\rm r} \sigma^4 f_r^2}{[\sigma^2+4(f_r-f_2)^2]^2}
&f_2\leq f_r<f_3,\\[2pt]
\end{cases}
\label{eq:omega_gw_1}
\end{equation}
where $E_{\rm gw}$ is the energy emitted in GWs, $M_{\rm chirp}\equiv
(m_1m_2)^{3/5}/(m_1+m_2)^{1/5}$ is the chirp mass, and $f_{i}$ ($i=1,
2, 3$) and $\sigma$ are frequencies that characterize the
inspiral-merger-ringdown waveforms, $\omega_{\rm m(r)}$ are normalization constants
chosen to make the waveform continuous,
and $\mathscr{F(G)_{\rm PN}}$ are the Post-Newtonian correction factors \citep{Ajith:2008by}.

We model the ensemble distribution of primary black hole masses as a power law
	\begin{equation}
	\label{eq:p-m1}
	p(m_1 |  \alpha, M_\mathrm{min}, M_\mathrm{max}) \propto
	\begin{cases}
	m_1^{-\alpha} & (M_\mathrm{min}\leq m_1 \leq M_\mathrm{max}) \\
	0 & (\mathrm{else})
	\end{cases}
	\end{equation}
and assume a flat distribution
	\begin{equation}
	\label{eq:p-m2}
	p(m_2 | m_1, M_\mathrm{min}) =
	\begin{cases}
	\frac{1}{m_1 - M_\mathrm{min}} & (M_\mathrm{min}\leq m_2\leq m_1) \\
	0 & (\mathrm{else})
	\end{cases}
	\end{equation}
of secondary masses. We set $ M_\mathrm{min}=5 \msun$ as the lower observed BH mass in low-mass X-ray binaries \citep{Farr:2011ct}. 
We set our fiducial values of $\alpha=1$, and $ M_\mathrm{max}=50 \msun$, however, we will explore the effect of varying these two parameters later in the paper. 
The upper limit on the BH mass is due to the assumption that pair-instability supernovae set an upper limit for the mass of the BHs born from stellar progenitors \citep{Woosley:2017dj}. 
The population analysis using BBH observations from LIGO/Virgo's first and second observing runs measured the spectral index
of the the primary BH power law to be $\alpha= 2.3^{1.3}_{-1.4}$ at 90\% confidence assuming a minimum black hole mass of 5$\msun$ and a maximum total mass of 100$\msun$ \citep{Abbottetal:2018vb}.
For the BNS systems we assume a uniform distribution between 1.3-1.5 $\msun$ for each of the NSs to be consistent with the range of observed masses of galactic double neutron star systems~\citep{Kiziltan:2013ky, Ozel:2016iv, Farrow:2019cl}, although we note that this choice is in tension with the masses of GW190425~\citep{Abbottetal:2020uq}.

\section{Results}

\begin{figure}
 \includegraphics[width=\columnwidth]{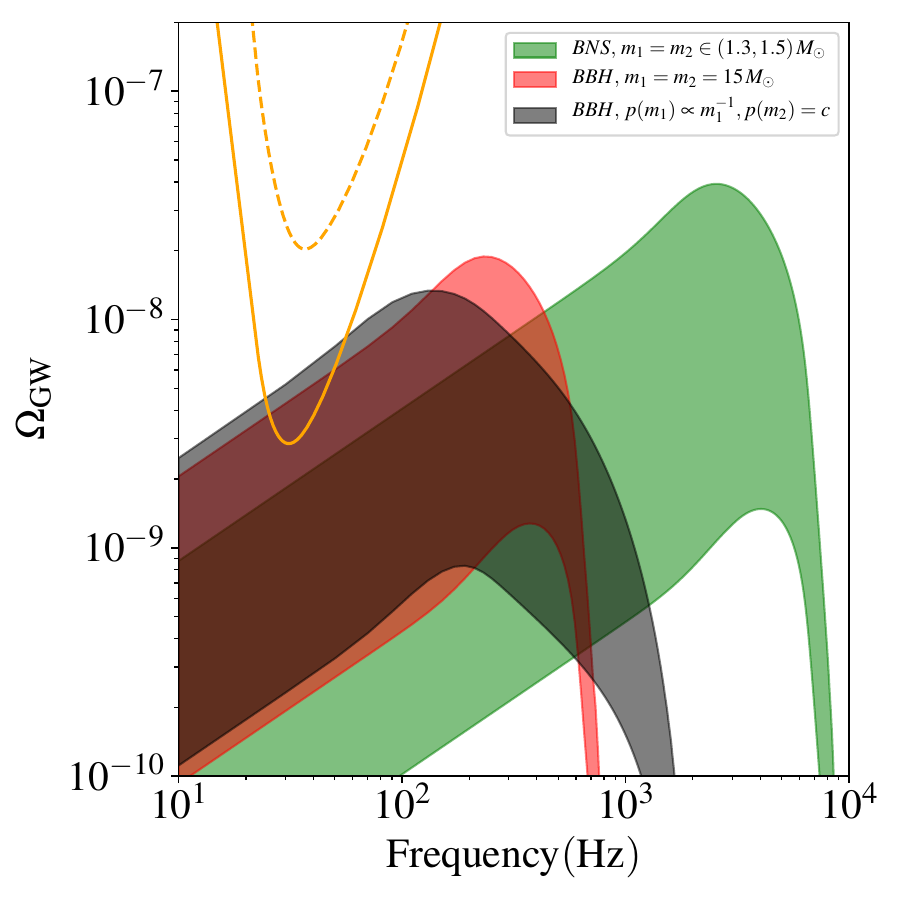}
 \caption{ The predicted SGWB incorporating the uncertainty due to both the local merger rate and the underlying DTD of the population of BBH and BNS mergers. 
 The green shaded region shows the predicted range of $\Omega_{\rm GW}$ for a population of BNS composed of two NSs each with a mass uniformly distributed between 1.3, and 1.5 $\msun$.
 For the BBH systems we considered two models: 
 (i) in which the black hole masses are drawn from a population where the primary BH mass follows $p(m_1)\propto m_1^{-1}$ bounded between 5 and 50 $\msun$, 
 and the secondary uniformly distributed between 5 and $m_1$. This scenario is shown with the black shaded region;  (ii) in which the two BHs each have a mass of 15 $\msun$, which is shown with the red shaded region. 
 The average chirp mass of the two BBH population models is the same.
 The solid orange line is the Power-law Integrated (PI) curve for adLIGO detection of the SGWB with $\mathrm{SNR}=3$ after one year of observing time. 
 The dashed orange line shows the $2-\sigma$ PI curve from LIGO's O1 and O2 runs~\citep{Abbottetal:2019kc}. 
 The cut off frequency is lower for the full population since the ISCO frequency of BHs is inversely proportional to their mass, and we have contributions from more massive BHs in the full population model. 
}
 \label{fig:omega_GW}
\end{figure}

\begin{figure}
 \includegraphics[width=\columnwidth]{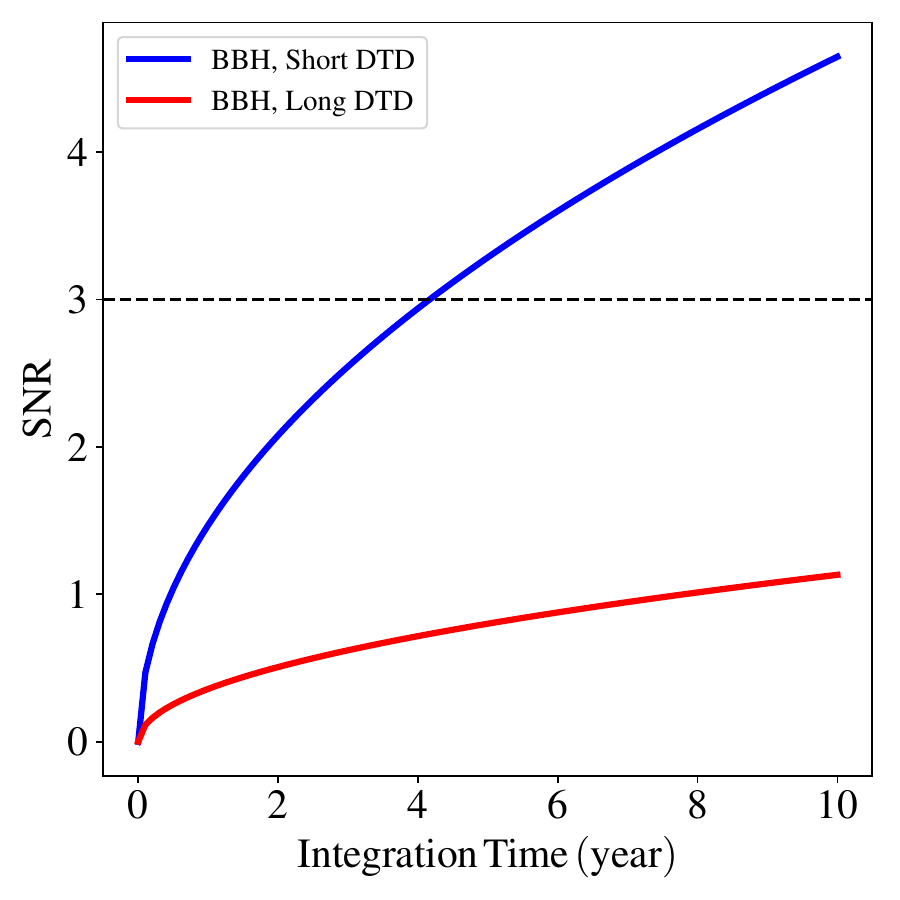}
 \caption{SNR for the detection of the SGWB from BBH mergers as a function of integration time assuming a local BBH merger rate of 30 $\rm Gpc^{-3} yr^{-1}$ and the fiducial power-law BH mass model, i). 
 The two curves show two extreme DTD models for the BBH mergers. 
 Within four years of observing time with adLIGO, one should be able to detect the SGWB with $\mathrm{SNR}> 3$ from BBHs for a short DTD. 
 Lack of detection can be used to rule out the existence of such DTDs for BBHs at fixed local BBH merger rate of 30 $\rm Gpc^{-3} yr^{-1}$.   }
 \label{fig:SNR}
\end{figure}

The results for the two extreme cases of DTDs for both the BBH and BNS systems are shown in Figure \ref{fig:omega_GW}. 
The green shaded region shows the predicted range of $\Omega_{\rm GW}$ for a population of BNS composed of two NSs each with a mass uniformly distributed between 1.3 and 1.5 $\msun$. 
The rate of the BNS mergers from LIGO is $\mathcal{R}^{\rm BNS}_0\approx 980^{+1490}_{-730}\rm  Gpc^{-3} yr^{-1}$ \citep{Abbottetal:2020uq}. 
In order to get the full possible range we assign the highest rate to the short DTD model, and 
the lowest rate to the long DTD model. We do the same for the BBHs where we consider a merger rate between 20 to 110 $\rm Gpc^{-3} yr^{-1}$~\citep{Abbottetal:2020uq}.
 For the BBH systems we considered two models in which: (i) the primary BH mass follows $p(m_1)\propto m_1^{-1}$ bounded between 5 and 50 $\msun$, 
 and the secondary is uniformly distributed between 5 $\msun$ and $m_1$ (grey shaded region)  (ii) each BH has a mass of 15 $\msun$ (red shaded region). 
 The average chirp mass of the two BBH population models are the same, $\langle M_{\mathrm{chirp}}\rangle \approx 13 \msun$, although their predicted $\Omega_{\rm GW}$ differs by a factor of 3 in magnitude. We also show the power-law integrated (PI) curve~\citep{Thrane:2013dy} for the detection of the SGWB with an SNR of 3 by adLIGO at design sensitivity~\citep{Abbott:2018gf} for one year of observing time and the $2-\sigma$ PI curve using data from LIGO's first and second observing runs, O1 and O2~\citep{Abbottetal:2019kc}.  Any background intersecting the PI curve will be detected with the specified significance within the given observing time. 

The SNR of SGWB detection can be computed following \citet{Allen:1997gg}:
\begin{equation}
{\rm SNR}\approx{3H_0^2\over 10\pi^2}\ \sqrt{T}\ 
\left[\int_{-\infty}^\infty df\
{\gamma^2 (|f|) \Omega_{\rm gw}^2(|f|) \over f^6 P_1(|f|) P_2(|f|)}
\right]^{1/2}\ .
\label{e:SNR_large_noise}
\end{equation}
Here $\gamma(f)$ is the overlap reduction function, which accounts for the separation and relative orientation of the detectors, a closed form of which is given in \citet{Flanagan:1993cw}. 
$P_1 (f)$ and $P_2 (f)$ are the noise power spectral densities of the detectors, and $T$ is the integration time. The results for the BBH mergers, assuming a local merger rate of 30 $ \rm Gpc^{-3}yr^{-1}$, are shown in Figure \ref{fig:SNR}. The curves represent two extreme DTD models for the BBH mergers. 
 Within four years of continuous observing time with adLIGO at design sensitivity, one should be able to detect the SGWB with $\mathrm{SNR}> 3$ from the BBHs if the DTD follows a fast-merging model. 
A lack of detection can rule out short DTDs for the BBHs. In this calculation we assumed that the BBHs follow case (i). 
The BBHs dominate over BNSs according to the results shown in the Figure \ref{fig:omega_GW} assuming $\rm \mathcal{R}^{\rm BBH}_0=30\,Gpc^{-3} yr^{-1}$ and  $\rm \mathcal{R}^{\rm BNS}_0=760\,Gpc^{-3} yr^{-1}$ in the local universe \citep{Abbottetal:2020uq}.

\begin{figure*}
 \includegraphics[width=\columnwidth]{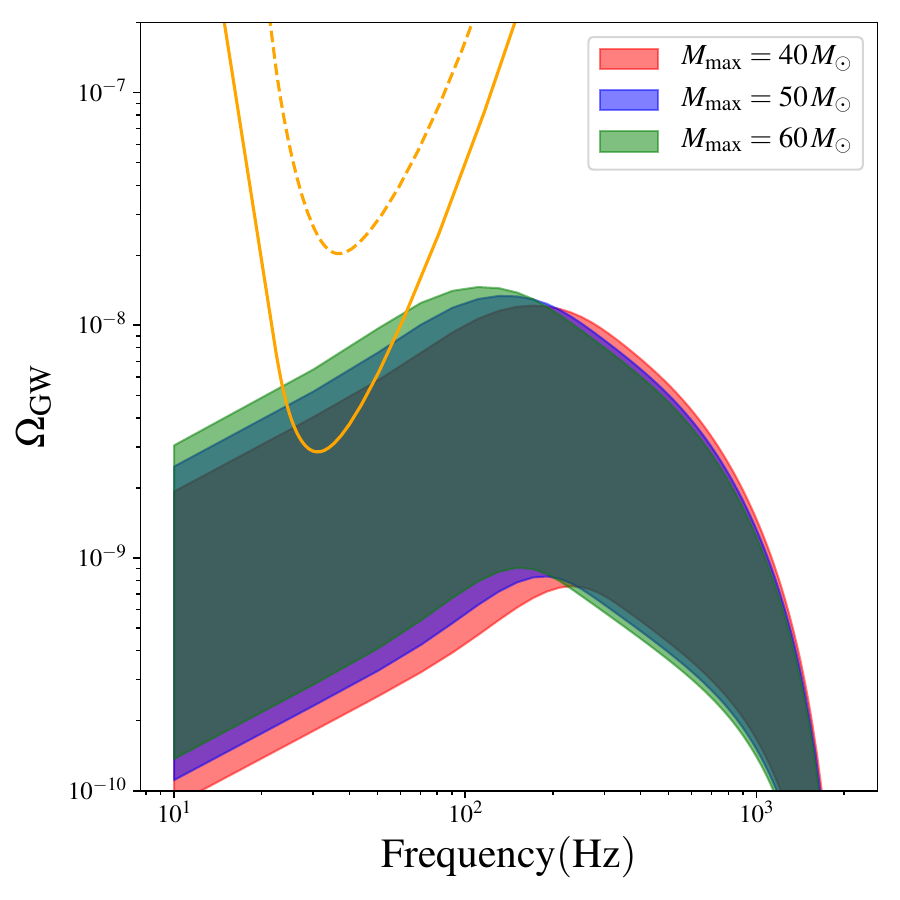}
  \includegraphics[width=\columnwidth]{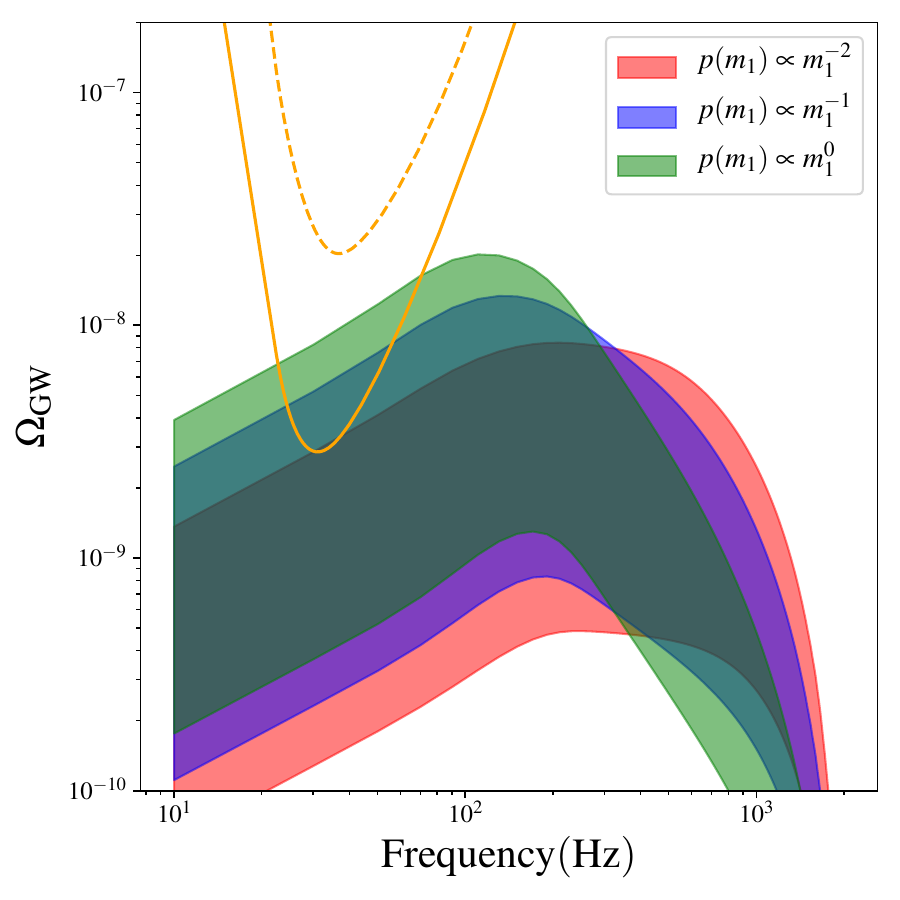}
 \caption{ \emph{Left panel:} The impact of variation in the maximum allowed BH mass, assuming $\alpha=1$, on the overall contribution to $\Omega_{\rm GW}$ from a population of BBHs.
 The shaded regions indicate the plausible range given both the DTD models and the local rate of the BBH mergers, assigning the lowest local rate to the longest DTD to obtain the lowest possible background, and assigning the highest local rate to the shortest DTD model to get the highest possible background level. 
 The orange line is the Power-law Integrated (PI) curve for the adLIGO detection of the SGWB with $\mathrm{SNR}=3$ after one year of observing time. 
 The dashed orange line shows the $2-\sigma$ PI curve from LIGO's O1 and O2 runs~\citep{Abbottetal:2019kc}. Abbottetal:2019kc
 \emph{Right panel:} The impact of variation in the slope of the primary black hole mass function ($\alpha$), fixing the 
 maximum mass to $M_{\rm max}=50\msun$. In both cases the uncertainty in DTD dominates over other uncertainty in $\alpha$ and $M_{\rm max}$ within the explored range in our work. 
}
 \label{fig:two_pararms}
\end{figure*}

So far our results have been based on assuming $p(m_1)\propto m_1^{-1}$, and $M_{\rm max}=50\msun$. However, these two assumptions on their own can impact the SGWB on top of the DTD assumptions~\citep{Talbot:2018cj, Jenkins:2019kq}. 
Here we relax both of these assumptions. In the left panel of Figure \ref{fig:two_pararms} we explore the impact of changing the maximum BH mass at a fixed value of $\alpha=1$. 
We see that a higher allowed maximum mass for the primary BH leads to a higher value of $\Omega_{\rm{GW}}$ while also shifting the cutoff frequency to lower values as more massive BHs have lower frequencies associated with their last stable circular orbit (ISCO). In the right panel we vary the exponent $\alpha$, while keeping the value of $M_{\rm max}=50 \msun$. 
In this case we only observe a rescaling of the curves with more positive slopes leading to a higher background level as expected. 
The shaded regions in all cases correspond to the different extreme DTD and local merger rate assumptions.  

Therefore the total uncertainty budget, assuming the local rate is constrained, comes from the DTD, the slope of the primary BH mass function, and the maximum allowed BH mass. 
However, we can conclude that the uncertainty in the DTD dominates over the uncertainty in $\alpha$ and $M_{\rm max}$ within the range explored in our work. 

\begin{figure*}
 \includegraphics[width=\columnwidth]{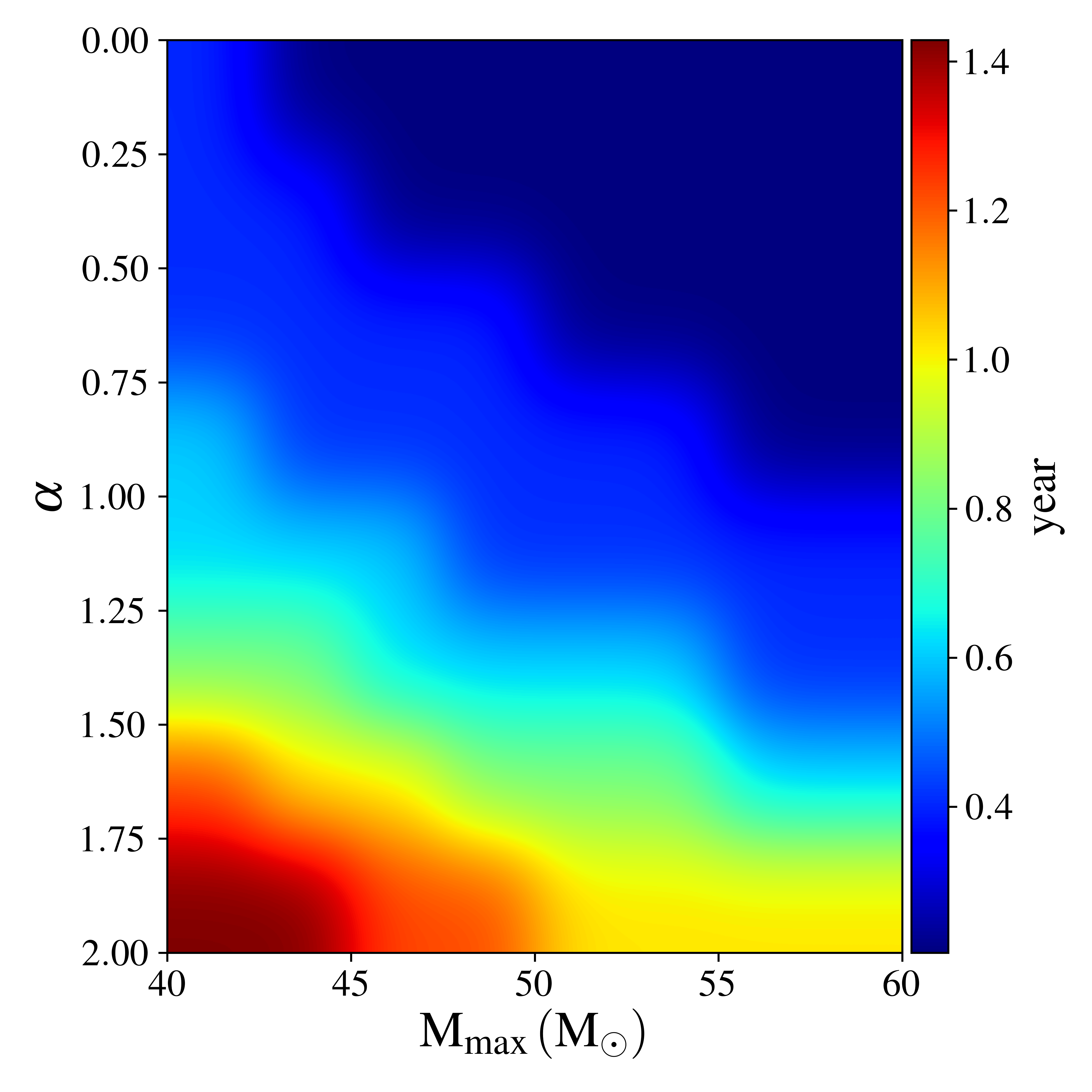}
  \includegraphics[width=\columnwidth]{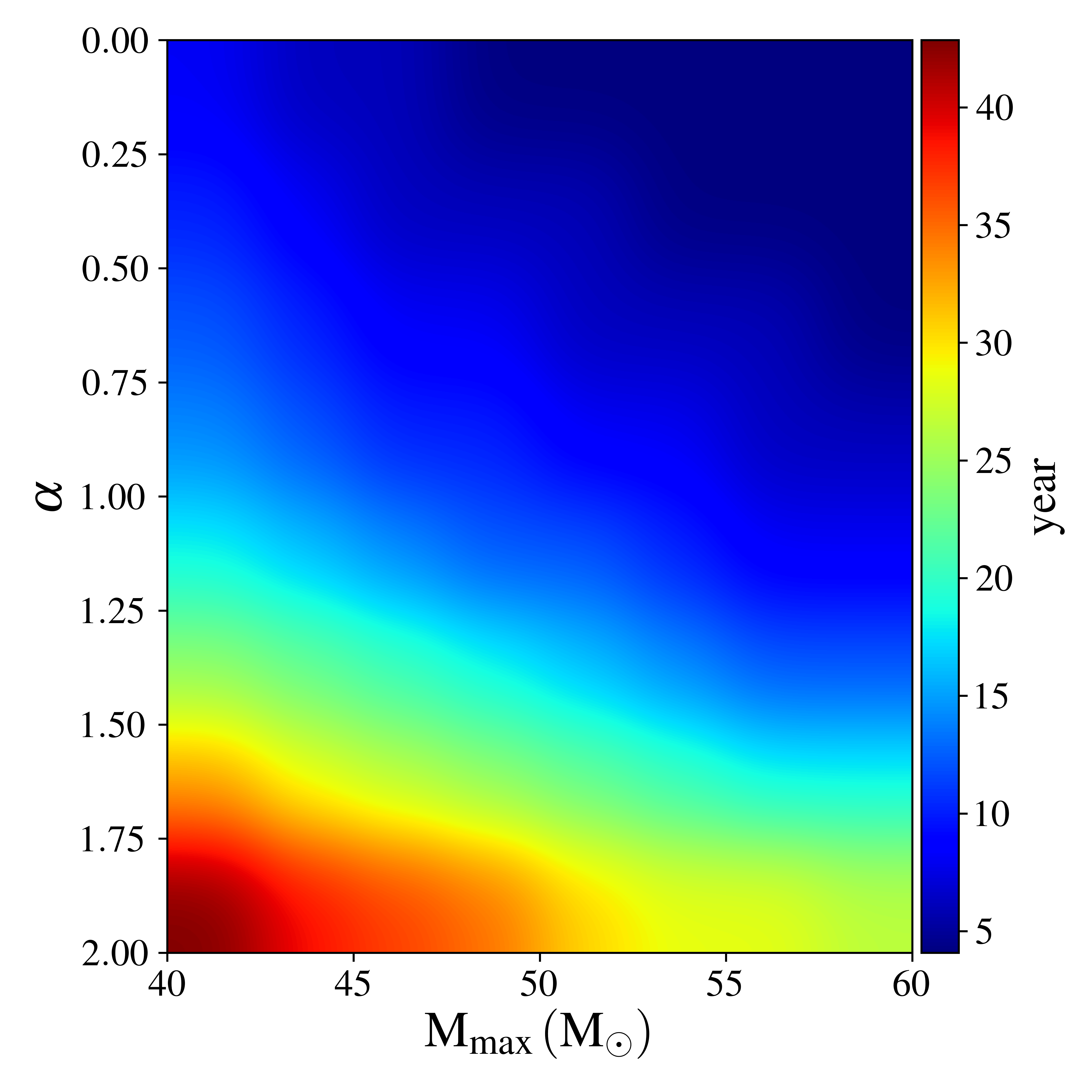}
 \caption{\emph{Left panel:} The integration time to detect the SGWB due to a BBH population with a local merger rate of 110 $ \rm Gpc^{-3}\,yr^{-1}$ with SNR>3 at a given $\alpha$ and $M_{\rm max}$. 
 The color coding shows the integration time in years 
 needed to detect such a background with SNR>3 assuming the short DTD. This plot shows that if the underlying distribution of the primary BH follows $p(m_1)\propto m_1^{-1}$, and $M_{\rm max}=50 \msun$, then
 with 0.5 yr of observing time the advanced LIGO detectors should be able to detect such background with SNR>3.
  \emph{Right panel:} assuming a local merger rate of 20 $ \rm Gpc^{-3}yr^{-1}$ and the long DTD. This plot shows that if the underlying distribution of the primary BH follows $p(m_1)\propto m_1^{-1}$, and $M_{\rm max}=50 \msun$, then
 with 10 yr of observing time the LIGO detectors should be able to detect such background with SNR>3.}
 \label{fig:t_obs}
\end{figure*}

The observing time at design sensitivity would depend on all these assumptions. The left panel of Figure \ref{fig:t_obs} shows the integration time needed to detect the SGWB assuming a short DTD, as a function of 
$\gamma$ and $M_{\rm max}$, for the advanced LIGO detectors operating at design sensitivity. The right panel shows the same but assuming a slow DTD. There are regions of the parameter space, such as a short DTD, large positive slope ($\gamma=1$), and large $M_{\rm max}=60 \msun$ that 
would need an integration time of less than half a year to be detected assuming an optimistic local merger rate of $110~\rm Gpc^{-3}\, yr^{-1}$, and therefore a non-detection of a background can rule out that part of the parameter space as the constraint on the local merger rate improves.

%

\begin{figure}
 \includegraphics[width=\columnwidth]{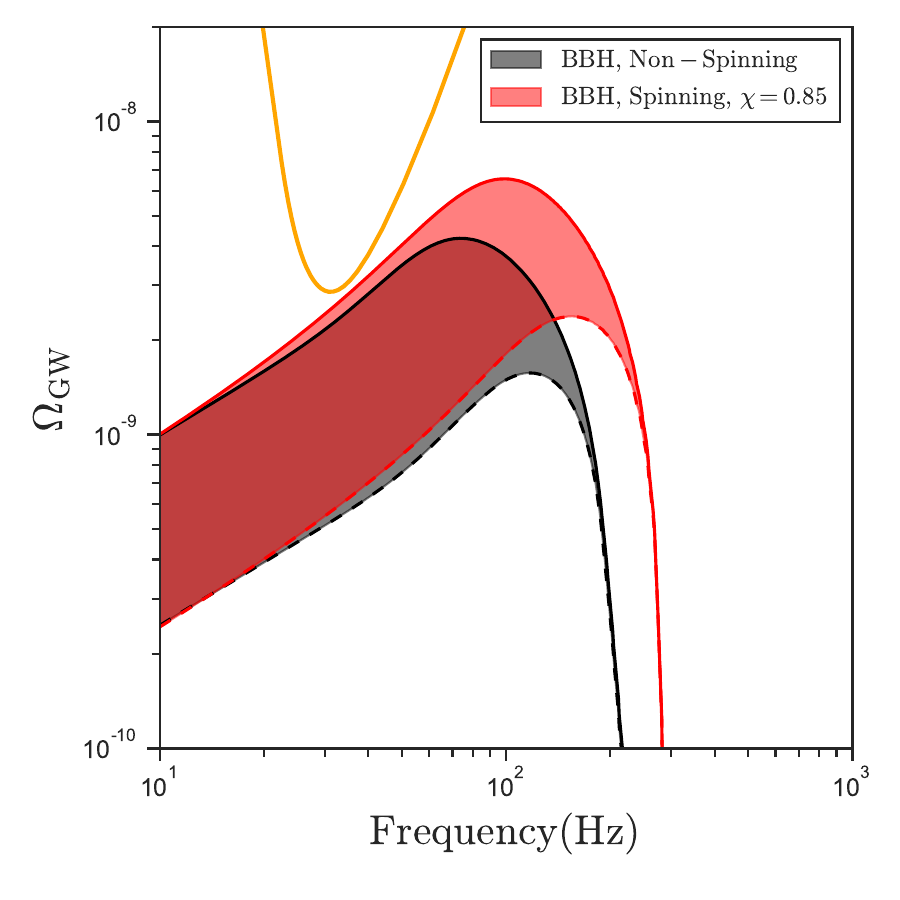}
 \caption{The impact of BH spin on the spectral energy density of the GW background. We have assumed a fixed local merger rate of 30 $\rm Gpc^{-3} yr^{-1}$, and the shaded regions show 
 the variation due to DTD assumptions. In the grey shaded region we show the case when the BHs are assumed to be non-spinning. 
 The red shaded region shows the result for the same population of the BBHs when all the BHs have $\chi=0.85$. 
 The impact is sub-dominant compared to the other parameters, such as $\alpha$, and $M_{\rm max}$. The solid orange line is the PI curve for adLIGO detection of the SGWB with $\mathrm{SNR}=3$ after one year of observing time. 
 }
 \label{fig:spin}
\end{figure}

We also study the impact of spin on the overall contribution to $\Omega_{\rm GW}$. For this purpose we construct $\frac{dE_{\rm gw}}{df_r}$ for non-precessing, spinning BBHs following \citep{Ajith:2011ha}.
Figure \ref{fig:spin} shows this effect, comparing a non-spinning population to one in which all the BHs have $\chi=0.85$ which is the maximum spin considered in the construction of the waveform models in  \citet{Ajith:2011ha}. 
Here we have assumed population ii) of equal mass BBH mergers at a local merger rate of $30~\mathrm{Gpc^{-3}\,yr^{-1}}$.
The overall impact is subdominant compared to the other parameters we considered in this work. 
The only noticeable difference is the increase of the cut off frequency which arises from the fact that the ISCO radius of a spinning BH is smaller compared to a non-spinning BH, and therefore its associated frequency is higher.

\section{Future GW detectors}

\begin{figure}
 \includegraphics[width=\columnwidth]{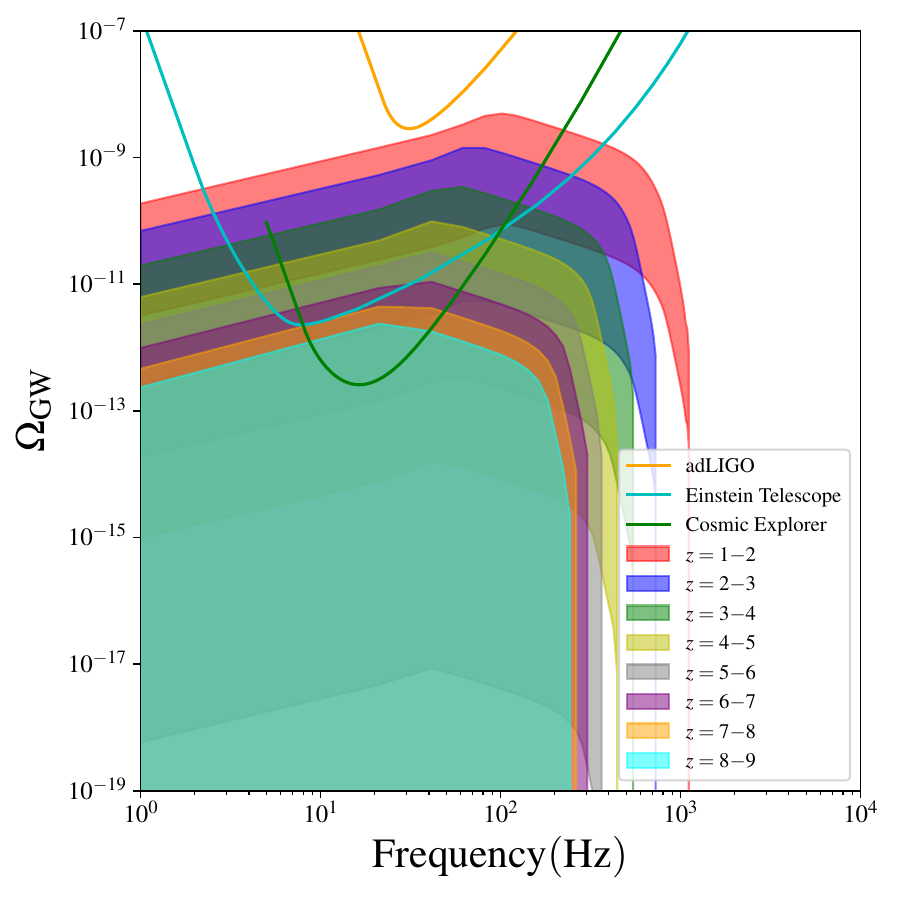}
 \includegraphics[width=\columnwidth]{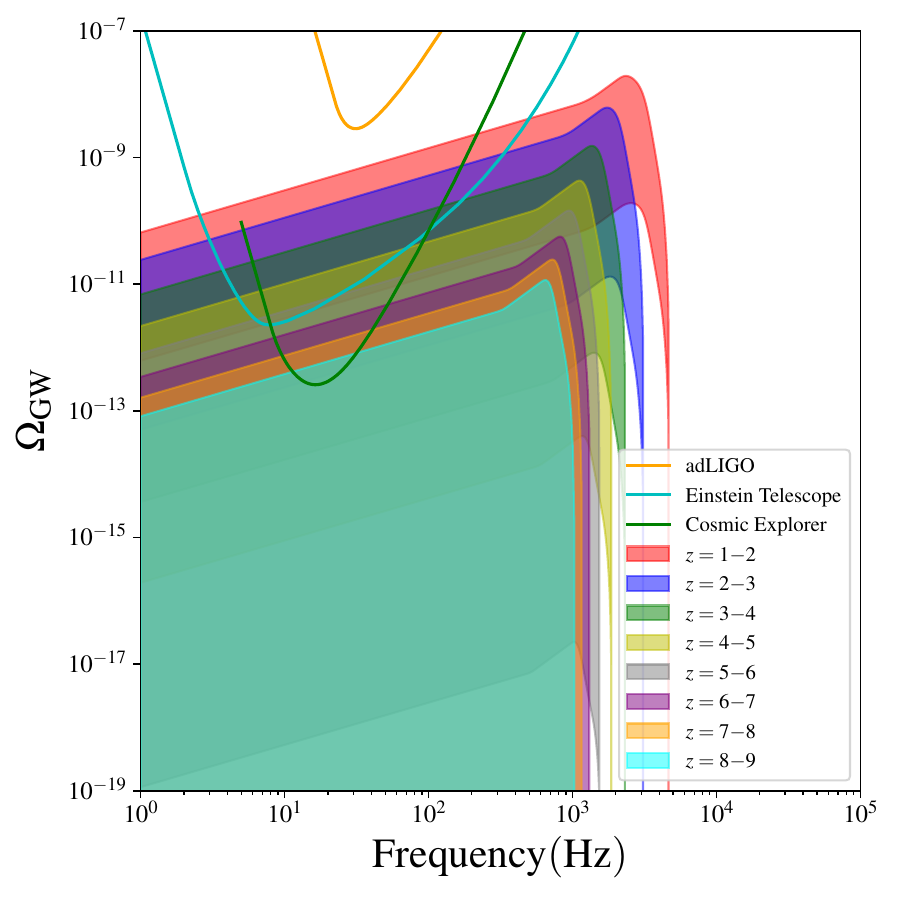}
 \caption{\emph{Top panel:} The predicted SGWB for BBHs in different redshift slices. Each shaded area indicates the estimated contribution to $\Omega_{\rm GW}$ of the BBHs in a specific redshift range. 
 We also show the PI curves for the adLIGO, ET, and CE detectors for an SNR of 3 within one year of observing time. \emph{Bottom panel:} The same but for the BNSs. Third generation gravitational-wave detectors, such as ET and CE, would be able to 
detect the contributing background from both the BBH and BNS populations at redshifts $z>6$; however, novel SGWB search techniques should be implemented to distinguish between the contributions from the two populations. 
 }
 \label{fig:z_dist}
\end{figure}

We separate the contribution of the BBHs at each redshift slice to the overall SGWB and show the result in the left panel of Figure \ref{fig:z_dist}. 
While the majority of the contribution originates from the BBHs at low redshifts, third generation GW detectors, such as Einstein Telescope~\citep{Punturo:2010jf} and Cosmic Explorer~\citep{Abbott:2017ie,Dwyer:2015df}, would be able to 
detect the contributing background from both the BBH and BNS populations at redshifts $z>6$. Because the BBH population is expected to be completely resolved across cosmic history by 3G detectors~\citep{Vitale:2019cy}, the properties of these sources can be extracted from individual detections, but this presents a computational challenge since conducting full source characterization using Bayesian inference for so many high-SNR events is costly~\citep{Vitale:2017cl}. The search for the SGWB from these sources, however, offers a computationally inexpensive alternative, as it relies on searching for excess correlated power in the detectors without modeling the individual sources using waveforms. Additionally for BNS mergers, the SGWB will serve as a probe of the significant fraction of the population that remains individually undetectable even with 3G detectors~\citep{Safarzadeh:2019kj}.

 We also show the PI curves for Cosmic Explorer and Einstein Telescope using the ET-D design configuration in Figure \ref{fig:z_dist}. We consider a network of two Cosmic Explorer instruments with the same overlap reduction function as the Hanford-Livignston detector pair. For the Einstein Telescope, we use the overlap reduction function for two V-shaped detectors separated by $120^{\circ}$ following \cite{Regimbau:2012if}. We note that the SGWB cannot be detected using a single 3G interferometer, as instrumental power cannot be separated from astrophysical power without cross-correlating between detectors with different noise sources. This presents a challenge for detecting the SGWB with ET alone, as even though it consists of three nested interferometers, they will share common instrumental and environmental noise sources.

While the overall background level is detectable by standard cross correlation techniques \citep{Allen:1997gg}, such techniques would not be able to single out the contribution from each of the CBO populations (BBHs vs BNS) separately. 
However, new approaches \citep{Smith:2018hw,Vivanco:2019ii,Smith:2020ie} have recently been proposed to search for the SGWB from BBHs and BNSs separately. These new searches use the fact that the astrophysical background is non-Gaussian--since individual compact binary systems do not merge simultaneously--and represent the background as a sum of individual sub-threshold detections modeled by waveforms. Because BBH and BNS systems lie in disparate parts of the mass parameter space, searches can be conducted for the background from each type of sources separately. The merger rate for each type of compact binary system inferred using these search methods can be converted into a stochastic background energy density, so the individual DTDs for BBH and BNS can be constrained using the framework described in this work. These non-Gaussian searches will serve as a hybrid, allowing both the individual detections and the sub-threshold population to be probed simultaneously when implemented for 3G detector sensitivity~\citep{Smith:2020ie}.

\begin{figure}
 \includegraphics[width=\columnwidth]{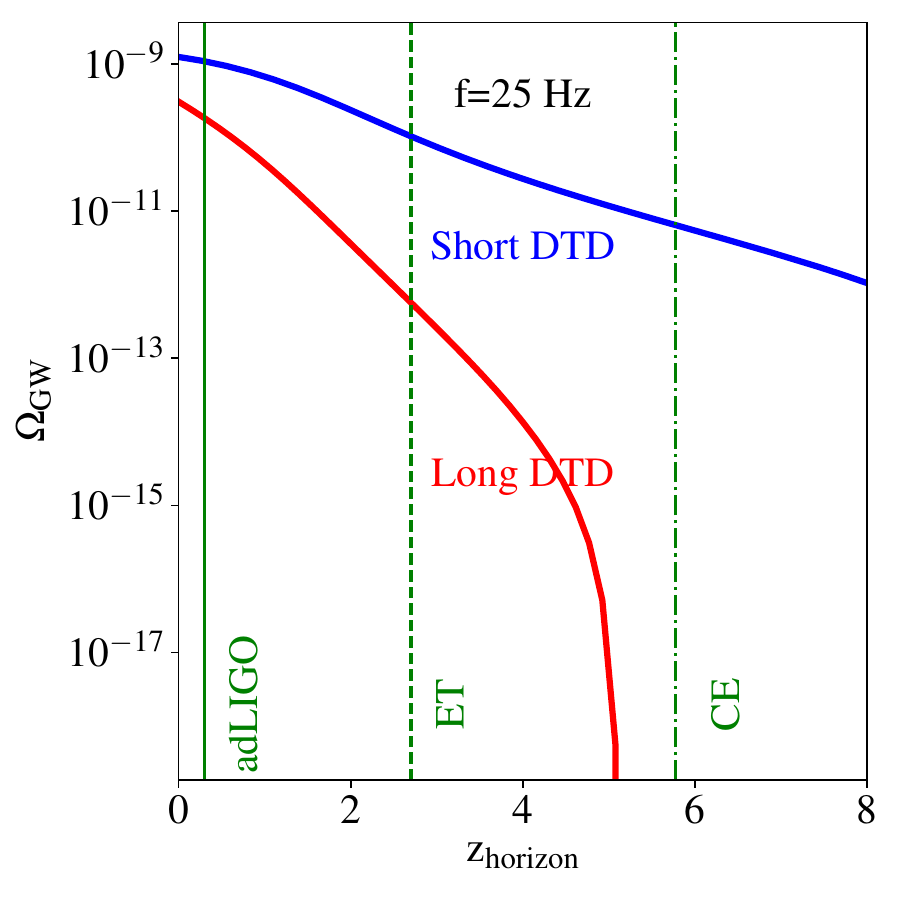}
 \caption{The predicted SGWB for BBHs at $f=25$ Hz as a function of detection horizon.
  As the detection horizon increases, the SGWB decreases.
 The effect is more drastic for long DTD (red line) than short DTD (blue line) because in the long DTD case the contribution to the SGWB is already limited to redshifts $z<3$, 
 and therefore increasing the detection horizon to such redshifts leaves no room for additional contributions to the background. 
 The vertical lines indicate the horizon distance for adLIGO, ET, and CE for the case of a population of BBHs with $m_1=m_2=15\msun$.}
 \label{fig:horizon}
\end{figure}

If we are interested in probing the sub-threshold population alone, after removing the individually detected sources, one can measure the background. 
Figure \ref{fig:horizon} shows the level of $\Omega_{\rm GW}$ as a function of detection horizon for the population of BBHs that we have considered in this work.
As the detection horizon increases, the expected contribution to the the background naturally drops. This is more pronounced for a long DTD as the contribution to the background is already limited to 
sources at $z<3$ and therefore increasing the detection horizon beyond that leaves little room for the SGWB contribution from the BNS or BBH systems. We note that constraints on the DTD can be used to inform the design of 3G detectors, since for longer delay times, there will be no astrophysical compact binary mergers at high redshifts. 

We also show the horizon redshift for advanced LIGO design sensitivity, Cosmic Explorer, and Einstein Telescope in Figure~\ref{fig:horizon}. The horizon distance is defined as the maximum distance at which a source would be detectable with an SNR above some threshold, $\rho > \rho_{\min}$, where the optimal SNR for a compact binary source with waveform $h(f)$ observed by a detector with power spectral density $P(f)$ is
\begin{align}
    \rho^{2} = 4\int_{0}^{\infty}\frac{h(f)^{*}h(f)}{P(f)}df. 
\end{align}
The waveform $h(f)$ includes the contribution from the detector antenna patterns, which account for the sensitivity of the detector to the two gravitational-wave polarizations due to its geometry. We follow \cite{Schutz:2011fn} and average the antenna patterns for the three interferometer configurations over the sky using the \texttt{bilby} software~\citep{Ashton:2019cn}. We compute the waveform for the BBH population ii) specified above for a source with an inclination angle of $28.6^{\circ}$, which is the most likely inclination angle for observed sources after accounting for selection biases~\citep{Schutz:2011fn}. Because the waveform is proportional to the inverse of the luminosity distance $d_{L}$ to the source, the horizon distance can be calculated as
\begin{align}
    d_{L, \max} = d_{0}\frac{\rho}{\rho_{\min}}
\end{align}
where $d_{0}$ is some reference distance used to calculate $\rho^{2}$. This can then be converted into a redshift assuming the cosmological parameters specified in \S2. We choose $\rho_{\min} = 8$ and consider only a single detector of each sensitivity, with ET consisting of three individual interferometers. For the source described above, we obtain $z_{\max}$ = 0.28, 2.73, and 5.77, for advanced LIGO, ET, and CE, respectively. We note that heavier sources with different orientations can be observed out to much higher redshifts for all three detectors, but use these values for representation.

\section{summary and discussion}

The distribution of the time lag between the birth and merger of compact binary objects due to the emission of gravitational waves (the delay time distribution, or DTD)
 can be characterized by a simple power law with two parameters, namely the minimum delay time and the slope of the power law. 
We examined two extreme cases of short and long DTDs, where in the fast version the time difference between the birth and merger is typically short compared to the 
age of the universe, and in the slow version it is comparable or longer. 
By normalizing to the observed rate of BBH or BNS mergers in the local universe, each of the DTD models results in a
vastly different population of undetected sources at high redshifts. 
These undetected populations will show themselves in the correlated searches for the stochastic background signal between different detectors. 

We studied the impact of the delay time distribution on the stochastic gravitational wave background (SGWB), 
and showed that the background level can change by a factor of $\approx5$ in the case of the BBHs assuming the two opposite ends of the DTDs. 
The difference of the background level is similar for the BNS systems. We note that the cut off frequency for the BNSs and BBHs are different. The cut off frequency is sensitive to the mass function we considered for the BBHs. 
The presence of more massive BHs in a population shifts the cut off frequency to lower values as the ISCO frequency of a BH scales inversely with its mass. 
In our fiducial model assuming a local BBH merger rate of 30 $\rm Gpc^{-3}yr^{-1}$, the level of the SGWB in the case of the short DTD is loud enough that it would be detectable after about four years of observing time with advanced LIGO at design sensitivity, and therefore a null detection can in principle rule out such DTD models.

We further explore the role of the primary BH mass function and the maximum allowed mass for the BHs on the contribution to the $\Omega_{\mathrm{GW}}$ and show that within the ranges of these parameters that we studied 
in our work, the uncertainties associated with these parameters is sub-dominant compared to the DTD variations. 

We further show that the next generation of GW detectors such as Einstein Telescope and Cosmic Explorer would be able to peer into the SGWB for sources above reionization redshifts for both  BNS and BBH mergers.
However, one has to rely on novel techniques to single out the contribution of the BBHs from the BNSs at such high redshift.

We have not included the same analysis for the BH-NS binaries as their merger rate is highly uncertain and expected to be less than the BBH mergers \citep{Dominik:2015dp}. We have also only considered a power-law parameterization for the DTD. Since we have shown that the unceranty in the amplitude of the SGWB is dominated by the uncertainty of the DTD, it will be interesting to consider the effect of other parameterizations in the future.
Similar studies to ours have been carried out in the literature before, although not with the main focus being to constrain the DTD of the CBOs \citep{Zhu:2011fu,Zhu:2013eh}.
Recently \citet{Callister:2020ua} put a constraint on the slope of the merger rate using the BBH detections and SGWB search results from the O1 and O2 LIGO observing runs. 

In their formulation, the evolution of merger rate density with redshift is constrained through the detection of the SGWB using a phenomenological model assuming that the stochastic background is due entirely to a population of compact binaries that is describable by a single mass distribution. In our work we assign the evolution of the merger rate with redshift to the underlying DTD of a given population. Moreover, our method allows for isolating the contributions from BBH and BNS mergers and determining the DTD for each type of compact binary independently, which is not captured by the framework presented in the \citet{Callister:2020ua} analysis. In addition, we show how 3G detectors can further constrain the shape of the DTDs.

A preference for short DTD through the SGWB searches would indicate that BBHs all form from a fast-merging channel irrespective of the metallicity at which their progenitor is born. 
This would go against the currently accepted paradigm of BBH assembly. 
On the other hand, a long DTD would indicate the local BBH merger events have originated from low metallicities at high redshifts and therefore provide us clues with respect to the conditions of BBH formation at such metallicities.

\acknowledgements

MTS is thankful to Enrico Ramirez-Ruiz, Evan Scannapieco, Matias Zaldarriaga, Tom Callister, Will Farr, Salvatore Vitale, and Edo Berger for useful discussions. 
MTS thanks the Heising-Simons Foundation, the Danish National Research Foundation (DNRF132) and NSF (AST-1911206 and AST-1852393) for support. 
This material is based upon work supported by the National Science Foundation under Grant No. AST-1440254. This work was supported in part by the Black Hole Initiative at Harvard University, which is funded by JTF and GBMF grants. 
SB. acknowledges support of the National Science Foundation, and the LIGO Laboratory. 
LIGO was constructed by the California Institute of Technology and Massachusetts Institute of Technology with funding from the National Science Foundation and operates under cooperative agreement PHY-1764464.
SB is also supported by the Paul and Daisy Soros Fellowship for New Americans and the NSF Graduate Research Fellowship under Grant No. DGE-1122374.

\bibliographystyle{yahapj}
\bibliography{the_entire_lib}

\end{document}